\documentclass[12pt]{article}
\usepackage{times}
\usepackage{amsmath,amssymb}
\usepackage{latexsym}
\newtheorem{assumption}{Assumption}[section]
\newtheorem{theorem}[assumption]{Theorem}

\newtheorem{lemma}[assumption]{Lemma}

\begin{document}
\title{Canonical analysis of space-time noncommutative theories
and gauge symmetries}
\author{Dmitri V. Vassilevich\thanks{Also at Department of Physics,
St.~Petersburg State University, Russia. 
}\\ {\it Institut f\"{u}r Theoretische Physik,
Universit\"{a}t Leipzig,}\\ {\it Augustusplatz 10, D-04109 Leipzig, Germany }
\\{email: \texttt{Dmitri.Vassilevich@itp.uni-leipzig.de}}}
\date{LU-ITP 2004/025}
\maketitle
\begin{abstract}
We construct a modification of the Poisson bracket which is suitable
for a canonical analysis of space-time noncommutative field theories.
We show that this bracket satisfies the Jacobi identities and
generates equations of motion. In this modified canonical formalism
one can define the notion of the
first-class constraints, demonstrate that they
generate gauge symmetries, and derive an explicit form of these
symmetry transformations. 
\end{abstract}
\section{Introduction}
Among all noncommutative field theories (cf. reviews \cite{NCrevs})
the theories with space-time noncommutativity have a somewhat lower
standing since it is believed that they cannot be properly quantised
because of the problems with causality and unitarity (see, e.g.,
\cite{problems}). Such problems occur due to the time-nonlocality
of these theories caused by the presence of an infinite number of
temporal derivatives in the Moyal star product. However, it has been
shown later, that unitarity can be restored \cite{unirest}
(see also \cite{Balachandran:2004rq})
in space-time noncommutative theories and that the path integral
quantization can be performed \cite{Fujikawa:2004rt}.
This progress suggests that space-time noncommutative theories
may be incorporated in general formalism of canonical quantization
\cite{cabooks}. Indeed, a canonical approach has been
suggested in \cite{Gomis:2000gy}.

Apart from quantization, there is another context in which canonical
approach is very useful. This is the canonical analysis of constraints
and corresponding gauge symmetries \cite{cabooks}. The problem of
symmetries becomes extremely complicated in noncommutative theories.
Already at the level of global symmetries one see phenomena which
never appear in the commutative theories. For example,
the energy-momentum tensor in translation-invariant noncommutative
theories is not locally conserved (cf. pedagogical comments in
\cite{Gerhold:2000ik}). At the same time all-order renormalizable
noncommutative $\phi^4$ theory is 
\emph{not} translation-invariant \cite{GrWu}.
A Lorentz-invariant interpretation of noncommutative space-time leads
to a twisted Poincare symmetry \cite{Chaichian:2004za}.
It is unclear how (and if) this global symmetry can be related to local
diffeomorphism transformations analysed, e.g., in \cite{Jackiw:2001jb}.
Proper deformation of gauge symmetries of generic two-dimensional
dilaton gravities remains on open problem (cf. 
footnote \ref{foot2d}).  
Solving (some of) the problems related to gauge symmetries 
in noncommutative field theories by the canonical methods is the main 
motivation for this work.

We start our analysis from the very beginning, i.e. with a definition of the
canonical bracket. Our approach is based on two main ideas.
First of all, we separate implicit time derivatives (which are
contained in the Moyal star), and explicit ones (which survive in
the commutative limit). Only explicit derivatives define the canonical
structure. As a consequence, the constraints and the hamiltonian
become non-local in time. Therefore, the notion of same-time
canonical brackets becomes meaningless. We simply postulate a
bracket between canonical variables taken at different
points of space ($\mathbf{x}$ and $\mathbf{x}'$) and of time
($t$ and $t'$): 
\begin{equation}
\{q_a(\mathbf{x},t),p^b(\mathbf{x}',t')\} =\delta_a^b \delta (\mathbf{x}-
\mathbf{x}')\delta (t-t')\label{stbra}
\end{equation}
This bracket is somewhat similar to the one appearing in
the Ostrogradski formalism for theories with
higher order time derivatives (see, e.g., \cite{Ostro} 
for applications to field theories
and  \cite{Gomis:2000gy} for the use in space-time noncommutative theories),
but there are important differences (a more detailed comparison
is postponed until sec.\ \ref{scon}).  

Of course, the proposed formalism means a departure
from the standard canonical procedure. Nevertheless, we are able to
demonstrate that the new bracket satisfies such fundamental requirements
as antisymmetry and the Jacobi identities. These brackets generate 
equations of motion. Moreover, one can define the notion of
first-class constraints with respect to the new bracket and
show that these constraints generate gauge symmetries of the action.
We shall derive an explicit form of the symmetry transformation
and see that they look very similar to the commutative case
(the only difference, in fact, is the modified bracket and the
star product everywhere).
%%%%%%%%
\section{Canonical bracket}
Consider a space-time manifold $\mathcal{M}$ of dimension $D$. Consider
the Moyal product of functions on $\mathcal{M}$ 
\begin{equation}
f\star g = f(x) \exp \left( \frac i2 \, \theta^{\mu\nu}
\overleftarrow{\partial}_\mu \overrightarrow{\partial}_\nu \right)
g(x) \,.\label{starprod}
\end{equation}
In this form the star product
has to be applied to plane waves and then extended
to all (square integrable) functions by means of the Fourier series.
$\theta$ is a constant antisymmetric matrix. We impose no restrictions on
$\theta$, i.e. we allow for the space-time noncommutativity.

The Moyal product is closed,
\begin{equation}
\int_{\mathcal{M}} d^Dx f\star  
g=\int_{\mathcal{M}} d^Dx f\times  g \label{clo}
\end{equation}
(where $\times$ denotes usual commutative product),
it respects the Leibniz rule
\begin{equation}
\partial_\mu (f\star g)=(\partial_\mu f)\star g + f\star (\partial_\mu g),
\label{Leib}
\end{equation}
and allows to make cyclic permutations under the integral
\begin{equation}
\int_{\mathcal{M}} d^Dx f\star  g \star h 
=\int_{\mathcal{M}} d^Dx h\star f\star  g
\label{cyper}
\end{equation}

The phase space on $\mathcal{M}$ consists of the variables
$r_j$ which can be subdivided into canonical pairs $q,p$ and other
variables $\alpha$ which do not have canonical partners (these will
play the role of Lagrange multipliers or of gauge parameters).
We define a bracket $(r_j,r_k)$ to be $\pm 1$ on the canonical pairs,
\begin{equation}
(q_a,p^b)=-(p^b,q_a)=\delta_a^b \label{rbr}
\end{equation}
and zero otherwise (e.g., $(\alpha,p)=(p^a,p^b)=0$).
With this definition the bracket (\ref{stbra}) reads:
$\{r_i(x),r_j(x')\}=(r_i,r_j)\delta (x-x')$. Note, that we are not going
to use brackets between two local expressions (see discussion below).

Now we can define canonical brackets between star-local functionals
on the phase space. We define 
the space of star-local \emph{expressions} as
a suitable closure of the space 
of free polynomials of the phase space variables $r_j$ and their
derivatives evaluated with the Moyal star. Such expressions integrated over 
$\mathcal{M}$ we call star-local \emph{functionals}. 

Locality plays no important role here, since after the closure one can
arrive at expressions with arbitrary number of explicit derivatives
(besides the ones present implicitly through the Moyal star).
It is important, that all expressions can be approximated with
only one type of the product (namely, the Moyal one), and no
mixed expressions with star and ordinary products appear.
One also has to define what does ``suitable closure'' actually mean,
i.e. to fix a topology on the space of the functionals.
This question is related to the restrictions which one imposes on
the phase space variables. For example, the bracket of two
admissible functionals (see (\ref{canobr}) below) should be
again an admissible functional. This implies that all integrands
are well-defined and all integrals are convergent. Stronger restrictions
on the phase space variables mean weaker restrictions on the
functionals, and vice versa. Such an analysis cannot be done
without saying some words about $\mathcal{M}$ (or about its'
compactness, at least)\footnote{Some restrictions on $\mathcal{M}$
follow already from the existence of the Moyal product,
which requires existence of a global coordinate system at least
in the noncommutative directions.}. 
We shall not attempt to do this analysis here
(postponing it to a future work). All statements made below are
true at least for $r\in C^\infty$ and for polynomial functionals
(no closure at all). 

Obviously, it is enough to define the bracket
on monomial functionals and extend it to the whole space by
the linearity. Generically, two such monomial functionals read:
\begin{equation}
R=\int d^Dx\, \partial_{\kappa_1} r_1\star 
\partial_{\kappa_2} r_2 \star \dots
\partial_{\kappa_n} r_n \,,\qquad
\tilde R=\int d^Dx\, \partial_{\tilde \kappa_1}  
\tilde r_1 \star \partial_{\tilde \kappa_2} \tilde r_2 \star\dots
\partial_{\tilde \kappa_m}\tilde r_m \label{RtR}
\end{equation}
$\kappa_j$ is a multi-index, $\partial_{\kappa_j}$ is a differential
operator of order $|\kappa_j|$.
The (modified) canonical bracket of two monomials is defined by the equation
\begin{eqnarray}
&&\{ R,\tilde R \}=
\sum_{i,j} \int d^Dx\, \partial_{\kappa_j} \left( \partial_{\kappa_{j+1}}
r_{j+1} \star \dots \partial_{\kappa_{j-1}}
r_{j-1} \right) (r_j,\tilde r_i) \nonumber\\
&&\qquad\qquad\qquad \star
\partial_{\tilde\kappa_i} \left( \partial_{\tilde \kappa_{i+1}}
\tilde r_{i+1}\star \dots \partial_{\tilde \kappa_{i-1}}
\tilde r_{i-1} \right) (-1)^{|\kappa_j|+|\tilde \kappa_i|}\,.
\label{canobr}
\end{eqnarray}
In other words, to calculate the bracket between two monomials one has
to (i) take all pairs $r_j$, $\tilde r_i$; (ii) use cyclic
permutations under the integrals to move $r_j$ to the last place,
and $\tilde r_i$ -- to the first; 
(iii) integrate by parts to remove derivatives
from $r_j$ and $\tilde r_i$; (iv) delete $r_j$ and $\tilde r_i$,
put the integrands one after the other connected by $\star$
and multiplied by $(r_j,\tilde r_i)$; (v) integrate over 
$\mathcal{M}$. Actually, this is exactly the procedure one uses
in usual commutative theories modulo ordering ambiguities following
from the noncommutativity. 

The following Theorem demonstrates that the operation we have
just defined gives indeed a Poisson structure on the space of star-local
functionals.  
\begin{theorem} \label{ThPoi}
Let $R$, $\tilde R$ and $\hat R$ be star-local
functionals on the phase space. Then\\
\textrm{(1)} $\{ R,\tilde R\}=-\{ \tilde R,R\}$ (antisymmetry),\\
\textrm{(2)} $\{ \{ R ,\tilde R\}, \hat R\}+
 \{ \{\hat R , R\}, \tilde R\}+\{ \{\tilde R ,\hat R\}, R\}=0 $
(Jacobi identity).
\end{theorem}
\noindent \textbf{Proof}. We start with noting that since we do
not specify the origin of the canonical variables, the time
coordinate does not play any significant role, and the statements
above (almost) follow from the standard analysis \cite{cabooks}.
However, it is instructive to present here a complete proof as
it shows that one do not need to rewrite the star product through
infinite series of derivatives (so that the $\star$ product indeed
plays a role of multiplication). Again, it is enough to study the
case when all functionals are monomial ones. Then the first assertion
follows from (\ref{canobr}) and $(r_j,r_k)=-(r_k,r_j)$. Let
\begin{equation}
\hat R=\int d^Dx\, \partial_{\hat \kappa_1} 
\hat r_1 \star \partial_{\hat \kappa_2} \hat r_2 \star \dots
\partial_{\hat \kappa_p}\hat r_p \,. \label{hatR}
\end{equation}
Consider $ \{ \{ R , \tilde R\}, \hat R\}$. The first of the brackets 
``uses up'' an $r_j$ and an $\tilde r_i$. The second bracket uses
a variable with hat and another variable either from $R$ or from
$\tilde R$. Consider first the terms in the repeated bracket 
which use twice some variables from $R$. All such terms combine into
the sum
\begin{eqnarray}
&&\sum_{i,k,j\ne l} (-1)^{|\tilde \kappa_i|+|\hat \kappa_k|}
(r_j,\tilde r_i)(r_l,\hat r_k) \int d^Dx\, 
\partial_{\kappa_{l+1}}r_{l+1}\star \dots \partial_{\kappa_{j-1}}r_{j-1}
\nonumber\\
&&\quad\star \partial_{\kappa_j+\tilde\kappa_i} \left(
\partial_{\tilde \kappa_{i+1}} \tilde r_{i+1}\star \dots
\partial_{\tilde \kappa_{i-1}} \tilde r_{i-1}\right)\star 
\partial_{\kappa_{j+1}}r_{j+1}\star \dots \partial_{\kappa_{l-1}}r_{l-1}
\nonumber\\
&&\quad\star\partial_{\kappa_l + \hat\kappa_k} \left(
\partial_{\hat \kappa_{k+1}} \hat r_{k+1}\star \dots
\partial_{\hat \kappa_{k-1}} \hat r_{k-1}\right)
\nonumber
\end{eqnarray}
This complicated expression is symmetric with respect to interchanging the roles
of the variables with hats and the variables with tilde. Therefore, it is
clear that the terms having two brackets with $r$ in
$\{ \{\hat R , R\}, \tilde R\}$ have exactly the same form as above but
with a minus sign. No such terms (with two brackets with $r$) may appear
in $\{ \{\tilde R ,\hat R\}, R\}$. Therefore, this kind of terms are 
cancelled in $\{ \{ R ,\tilde R\}, \hat R\}+
 \{ \{\hat R , R\}, \tilde R\}+\{ \{\tilde R ,\hat R\}, R\}$. By repeating 
the same arguments for $\hat r$ and $\tilde r$ one proves
our second assertion. $\Box$

One can define a canonical bracket between functionals and densities
(star-local expressions) by the equation:
\begin{equation}
\{ R, h(r )(x) \} :=\frac{\delta}{\delta \beta (x)}
\left\{ R, \int d^Dy\, \beta (y)\star h(r)(y) \right\} \,.\label{gloloc}
\end{equation}
To construct brackets between two densities (i.e., to give a proper
extension of (\ref{stbra}) to nonlinear functions) one has to
define star-products with delta-functions which may be a very non-trivial
task. We shall never use brackets between densities.

To use the canonical bracket in computations of variation we need the
following technical Lemma.
\begin{lemma}\label{varlemma}
Let $p^a$ and $q_b$ depend smoothly on a parameter $\tau$.
We assume that the variables $\alpha (x)$ (these are the ones which do
not have canonical conjugates) do not depend on $\tau$. Let $h(r(\tau))$
be a star-local expression on the phase space. Then
\begin{eqnarray}
&&\partial_\tau \int d^Dx \beta \star h(r(\tau))=
\int d^D x\, \left( (\partial_\tau q_a)\star 
\left\{  \int d^Dy \beta \star h(r), p^a(x) \right\}\right.\nonumber\\
&&\qquad\qquad\qquad \left.- (\partial_\tau p^a)\star 
\left\{  \int d^Dy \beta \star h(r), q_a(x) \right\} \right)\label{l1}
\end{eqnarray}
\end{lemma} 
\noindent\textbf{Proof.} Obviously, it is enough to prove this Lemma
for $\beta=1$. Let us consider first the case when just one of the
canonical variables (say, $p^b$ for a just single value of $b$) 
depends on $\tau$, and when 
$h(r)=h_1(r)\star \partial_\kappa p^b\star h_2(r)$
where neither $h_1$ nor $h_2$ depend on $p^b$. Then
\begin{equation}
\partial_\tau \int d^Dx\, h(r)=
\int d^Dx\, h_1\star \partial_\kappa (\partial_\tau p^b)\star h_2
=(-1)^{|\kappa|} \int d^Dx\, \partial_\kappa (h_2\star h_1) \star
\partial_\tau p^b\,.
\label{l111}
\end{equation}
On the other hand, by using (\ref{canobr}), one obtains
\begin{equation}
\left\{ \int d^Dx\, h(r),\int d^Dy\, \beta(y)\star q_b (y) \right\}=
-(-1)^{|\kappa|} \int d^Dx\, \partial_\kappa (h_2\star h_1)\star \beta \,.
\label{l112}
\end{equation}
Next we use (\ref{gloloc}) to see that the statement of this Lemma is
indeed true for the simplified case considered. In general case
one has to sum up many individual contributions to both sides of (\ref{l1})
from different canonical
variables occupying various places in $h$.
Each of this contributions can be treated in the same way as above. $\Box$ 

As an application, consider a noncommutative field theory described by
the action
\begin{equation}
S=\int \left( p^a\partial_t q_a -h(p,q,\lambda ) \right) d^Dx
=\int p^a\partial_t q_ad^Dx -H \,,\label{Sh}
\end{equation}
where $h$ is a star-local expression,
it contains temporal derivatives 
only implicitly, i.e. only though the Moyal star. 
Note, that due to (\ref{clo})
the star between $p^a$ and $\partial_tq_a$ can be omitted.
If one takes into
account explicit time derivatives only, one can write
$p^a=\delta S/(\delta \partial_t q_a)$. Then, $H=S-\int p\partial_t q d^Dx$. 

The equations of motion generated from the action (\ref{Sh}) by
taking variations with respect to $q$ and $p$ can be written in
the ``canonical'' form:
\begin{equation}
\partial_t p^a + \{ H,p^a\}=0\,,\qquad \partial_tq_a +\{ H,q_a\}=0
\label{eom}
\end{equation}
This can be easily shown by taking $q(\tau)=q+\tau \delta q$ and
$p(\tau) = p+\tau \delta p$ and using Lemma \ref{varlemma}. No explicit
time derivative acts on $\lambda$. In a commutative theory $\lambda$
generates constraints.
%%%%%%
\section{Constraints and gauge symmetries}
Let us specify the form of (\ref{Sh}):
\begin{equation}
S=\int \left( p^a\partial_t q_a -\lambda^j \star G_j (p,q) 
-h(p,q) \right) d^Dx
\label{SwithG}
\end{equation}
We shall call $G_j(p,q)$ a constraint, although due to the
presence of the Moyal star it cannot be interpreted as a condition
on a space-like surface. 
Dirac classification of the constraints can be also performed with the
modified canonical bracket. We say that the constraints $G_j(p,q)$ are
first-class if their brackets with $h(p,q)$ and between each other
are again constraints, i.e.,
\begin{eqnarray}
&& \left\{ \int d^Dx \alpha^i\star G_i, \int d^Dx \beta^j\star G_j
 \right\} =\int d^Dx C(p,q;\alpha,\beta)^k \star G_k \,,\label{strc}\\
&& \left\{ \int d^Dx \alpha^i \star G_i, \int d^Dx h(p,q) \right\}
=\int d^Dx B(p,q;\alpha)^k \star G_k \,.\label{withh}
\end{eqnarray} 
By Theorem \ref{ThPoi}(1) the structure functions are antisymmetric,
$C(p,q;\alpha,\beta)^j=-C(p,q;\beta,\alpha)^j$. Further restrictions
on $C$ and $B$ follow from the Jacobi identities (cf. 
Theorem \ref{ThPoi}(2)).

\begin{theorem}\label{Tgau}Let $G_i(p,q)$ be fist-class constraints
(so that (\ref{strc}) and (\ref{withh}) are satisfied). Then the 
transformations
\begin{eqnarray}
&&\delta p^a = \left\{ \int d^Dx \alpha^j\star G_j,\, p^a\right\}\label{tr1}\\
&&\delta q_b = \left\{ \int d^Dx \alpha^j\star G_j,\, q_b\right\}\label{tr2}\\
&&\delta \lambda^j=-\partial_t \alpha^j -C(p,q;\alpha,\lambda)^j
-B(p,q;\alpha)^j\label{tr3}   
\end{eqnarray}
with arbitrary $\alpha^j$ are gauge symmetries of the action (\ref{SwithG}).
\end{theorem}
\noindent\textbf{Proof.} To prove this Theorem we simply check 
invariance of (\ref{SwithG}) under (\ref{tr1}) - (\ref{tr3}).
Let $f(p,q)$ be an arbitrary star-local
expression depending on the canonical variables $p$ and $q$ only.
Then, by (\ref{tr1}) and (\ref{tr2}),
\begin{equation}
\delta\, f(p,q)
=\left\{ \int d^Dx \alpha^j \star G_j,\, f(p,q)\right\}\label{tr4}
\end{equation}
It is now obvious that that the transformations of $G$ and $h$
in the action (\ref{SwithG}) are compensated by second and third
terms in $\delta\lambda$ respectively. The remaining term in the
action transforms as
\begin{eqnarray}
&&\delta \int d^Dx \, p^a\partial_t q_a=\nonumber\\
&&=
\int d^Dx \left( \left\{ \int d^Dy \alpha^j\star G_j,p^a(x)\right\} \star
\partial_tq_a
+ p^a \star\partial_t\left\{ \int d^Dy \alpha^j\star G_j,q_a(x)\right\}
\right)\nonumber\\
&&= 
\int d^Dx \left( \left\{ \int d^Dy \alpha^j \star G_j,p^a(x)\right\} \star
\partial_tq_a
-(\partial_t p^a)\star \left\{ \int d^Dy \alpha^j \star
 G_j,q_a(x)\right\}\right)
\nonumber\\
&&=\int d^Dx \alpha^j\star \partial_t G_j=
-\int d^Dx\partial_t( \alpha^j)\star G_j
\label{varvar}
\end{eqnarray}
Here we used integration by parts and Lemma \ref{varlemma}.
The last term in (\ref{varvar}) is compensated by the first (gradient)
term in the variation (\ref{tr3}). Therefore, the action (\ref{SwithG})
is indeed invariant under (\ref{tr1}) - (\ref{tr3}).$\Box$

\noindent\textbf{Example.} Consider a two-dimensional topological
noncommutative theory described by the action
\begin{equation}
S=\int \mathrm{tr}\left( \Phi \star F \right)
=\int d^2x \, \mathrm{tr} \left( \Phi \star F_{01}\right) \label{2dact}
\end{equation}
$F$ is a field-strength two-form with the components:
$F_{01}=\partial_0 A_1 -\partial_1 A_0 + [A_0,A_1]$. 
All commutators are taken with the Moyal star.
The scalar field $\Phi$ and the vector potential $A_\mu$
are matrix-valued. They can be expanded over generators the $T_J$
of a Lie group taken in an appropriate representation,
$\Phi=\Phi^J T_J$, $A_\mu=A_\mu^JT_J$. If the gauge group is
$U(1,1)$ this model is equivalent to a noncommutative version
\cite{Cacciatori:2002ib} of the Jackiw-Teitelboim gravity
\cite{JT}\footnote{\label{foot2d}
This model is the only two-dimensional dilaton
gravity so far which has a noncommutative counterpart. 
It is an interesting and important task to
deform other 2D gravities \cite{Grumiller:2002nm} as well.
Constructing and appropriate classical action with right number
of gauge symmetries seems to be the most hard part of the problem.
Quantization goes then \cite{Vassilevich:2004ym} as in the 
commutative case \cite{Kummer:1996hy}, 
at least for the model \cite{Cacciatori:2002ib}.}.
We suppose that the quadratic form $\eta_{IJ}=-\mathrm{tr}(T_IT_J)$
is non-degenerate, and that the algebra is closed with respect
to commutators and anti-commutators:
\begin{equation}
[T_I,T_J]={f_{IJ}}^KT_K,\qquad [T_I,T_J]_+=T_IT_J+T_IT_J=
{d_{IJ}}^KT_K \,.\label{stuc}
\end{equation}
The action (\ref{2dact}) can be rewritten as
\begin{eqnarray}
&&S=\int d^2x \left( -\eta_{IJ}\Phi^I\star \partial_0 A_1^J -A_0^I\eta_{IJ}
\star 
\left( \partial_1\Phi^J +\frac 12 {d_{KL}}^J [A_1^K,\Phi^L] \right.\right.
\nonumber\\
&&\qquad\qquad\qquad\qquad \left.\left. 
 +\frac 12 {f_{KL}}^J [A_1^K,\Phi^L]_+ \right)\right)\label{c2d}
\end{eqnarray}
According to this action, the canonical variables are $q^I=A_i^I$ and
$p_I=-\eta_{IJ}\Phi^J$. Lagrange multipliers $A_0^I$ generate the
constraints:
\begin{equation}
G_I=\eta_{IJ}
\left( \partial_1\Phi^J +\frac 12 {d_{KL}}^J [A_1^K,\Phi^L]
+\frac 12 {f_{KL}}^J [A_1^K,\Phi^L]_+ \right) \,.\label{2dcons}
\end{equation}
The structure functions
\begin{equation}
C(\alpha,\beta)^L=-\frac 12 \left( {d_{IJ}}^L [\alpha^I,\beta^J]
+{f_{IJ}}^L [\alpha^I,\beta^J]_+\right) \label{2dsc}
\end{equation}
do not depend on the canonical variables. The gauge transformations
(cf Theorem \ref{Tgau})
\begin{eqnarray}
&&\delta \Phi^J=-\frac 12 {d_{LI}}^J[\Phi^L,\alpha^I]
-\frac 12 {f_{LI}}^J[\Phi^L,\alpha^I]_+\label{de1}\\
&&\delta A_\mu^J =-\partial_\mu \alpha^J
-\frac 12 {d_{LI}}^J[A_\mu^L,\alpha^I]
-\frac 12 {f_{LI}}^J[A_\mu^L,\alpha^I]_+\label{de2}
\end{eqnarray}
are just infinitesimal versions of usual noncommutative gauge
transformations $\Phi \to e^{\alpha T}\star \Phi \star e^{-\alpha T}$,
$A_\mu \to e^{\alpha T}\star\partial_\mu  e^{-\alpha T}+
e^{\alpha T}\star A_\mu \star  e^{-\alpha T}$. The exponentials are
calculated with the star-product.

In a space-space noncommutative theory similar calculations were
done in \cite{Banerjee:2002qh}. 
%%%%%
\section{Conclusions}\label{scon}
In this paper we have suggested a modification of the Poisson
bracket which is defined on fields at different values of the
time coordinate. In this modified canonical formalism, only
explicit time derivatives (i.e., the ones which are not hidden
in the Moyal multiplication) define the canonical structure.
Although this means serious deviations from standard canonical
methods, the resulting brackets still satisfy the Jacobi identities
(Theorem \ref{ThPoi}) and generate classical equations of motion.
Our main result (Theorem \ref{Tgau}) is that we can still define
the notion of fist-class constraints, which generate gauge symmetries,
and these symmetries are written down explicitly\footnote{Just existence
of the symmetries does not come as a great surprise in the view
of the analysis of \cite{GT} which is valid for theories with
arbitrary (but finite!) order of time derivatives. 
An important feature of the present approach is rather simple
explicit formulae similar to that in the case of commutative theories
with 1st order time derivatives.}.

Let us compare the technique developed here to the Ostrogradski
formalism for theories with higher time derivatives.
In this formalism \cite{Ostro,Gomis:2000gy}
new phase space variables $P(t,T)=p(t+T)$ and 
$Q(t,T)=q(t+T)$ are introduced. Then $t$ is interpreted as an evolution
parameter, while $T$ labels degrees of freedom (number of degrees of freedom
is proportional to the order of temporal derivatives). Then a delta-function
$\delta (T-T')$ appears naturally on the right hand side of the Poisson
brackets between $Q$ and $P$ calculated at the same value of $t$.
By returning (naively) to the original variables $q$ and $p$ one obtains 
(\ref{stbra}). In the approach of \cite{Gomis:2000gy} one proceeds in
a different way. The resulting dynamical system is interpreted as
a system with an infinite number of second-class constraints.
Additional first-class constraints would lead to considerable complications
in this procedure. It may happen that these two approaches are
equivalent, but this requires further studies.

As for the prospects of the formalism developed in the present paper
one should mention first of all applications to particular physical
systems briefly outlined above. A more formal development would be
to construct a classical BRST formalism starting with our brackets.
Anyway, it is important to restore the reputation of space-time
noncommutative theories. This is required by the principles of
symmetry between space and time, but also by interesting 
physical phenomena which appear due to the space time 
noncommutativity (just as an example we may mention creation of
bound states with hadron-like spectra \cite{Vassilevich:2003he}).
%%%%%
\section*{Acknowledgments}
I am grateful to D.~M.~Gitman, D.~Grumiller, and P.~M.~Lavrov for
fruitful discussions.
This work was supported in part by the DFG Project BO 1112/12-1 and by
the Multilateral Research Project "Quantum gravity, cosmology and
categorification" of the Austrian Academy of Sciences and the
National Academy of Sciences of the Ukraine.
%%%%%

\end{document}